# Highly Efficient Ultrathin Light Emitting Diodes based on Perovskite Nanocrystals

Qun Wan[a,§], Weilin Zheng[a,§], Chen Zou[b,§], Francesco Carulli[c] Congyang Zhang[b], Haili Song[d], Mingming Liu[a], Qinggang Zhang[a], Lih Y. Lin[b], Long Kong[a], Liang Li[a,d*] and Sergio Brovelli[c*]

[a]School of Environmental Science and Engineering, Shanghai Jiao Tong University, 800 Dongchuan Road, 200240 Shanghai, P.R. China

[b]Department of Electrical and Computer Engineering, University of Washington, Seattle, WA 98195, U.S.A

[c]Dipartimento di Scienza dei Materiali, Università degli Studi di Milano-Bicocca, 20125 Milano, Italy

[d]Key Laboratory of Bioinorganic and Synthetic Chemistry of Ministry of Education, School of Chemistry, Sun Yat-Sen University, Guangzhou 510275, China

**Light-emitting diodes based on perovskite nanocrystals (PNCs-LEDs) have gained great interest for next-generation display and lighting technologies prized for their color purity, high brightness and luminous efficiency approaching the intrinsic limit imposed by extraction of electroluminescence from the device structure. Although the time is ripe for the development of effective light outcoupling strategies to further boost the device performance, this technologically relevant aspect of PNC-LEDs is still without a definitive solution. Here, following theoretical guidelines and without the integration of complex photonic structures, we realize stable PNC-LEDs with *EQE* as high as 29.2% (average *EQE*=24.7%), which substantially break the outcoupling limit of common PNC-LEDs and systematically surpass any previous perovskite-based device. Key to such unprecedented performance is channeling the recombination zone in PNC emissive layers as thin as 10 nm, which we achieve by finely balancing the electron and hole transport using $CsPbBr_3$ PNCs resurfaced with a nickel oxide layer. The ultra-thin approach general and, in principle, applicable to other perovskite nanostructures for fabricating highly efficient, color tunable transparent LEDs ideal for unobtrusive screens and displays and is compatible with the integration of photonic components for further enhanced performance.**

**Keywords:** Lead halide perovskite nanocrystals, light-emitting diodes, external quantum efficiency, light outcoupling, transparent LEDs

In the last few years, prized for their advantageous optical properties and affordable solution processability, lead halide perovskite nanocrystals (PNCs) have emerged as promising active materials in a wide range of photonic and optoelectronic technologies, spanning from photovoltaic cells[1-5], lasers[6,7] and radiation detectors[8,9] to luminescent solar concentrators[10,11] and artificial light sources[12-22]. Light emitting diodes based on PNCs (PNC-LEDs)[23-31] have experienced a particularly steep growth, owing to the efficient, spectrally tunable, narrow luminescence of PNCs and their defect tolerant electronic structure[24,32-37]. Building upon these uniqueness of PNCs, extensive research in material design, morphology control, surface chemistry,[13,15,24,28,34-36,38-41] and energy level engineering (via doping or alloying)[35,42] has been dedicated to understand and suppress detrimental surface defects - acting both as traps for electrically-injected carriers and as nonradiative quenching centers for the resulting excitonic luminescence - and to devise suitable molecular ligands and solution-based protocols for fabricating high quality, low-resistance PNC active layers[15,18,33,34,43]. These efforts have enabled researchers to incorporate PNCs with ~100% emission quantum yield ($\Phi_{PL}$) in devices featuring near unity carrier mobility ratio ($\gamma$), resulting in internal quantum efficiency ($IQE = \gamma \times \eta_{S/T} \times \Phi_{PL}$, corresponding to the ratio between the number of photons emitted from the active region and the number of electrons injected into the LED) close to 100% ($\eta_{S/T}$ is the singlet to triplet ratio that, in halide perovskite materials is close to one[44]). As a result, external quantum efficiency ($EQE=IQE\times\eta_{OC}$, corresponding to the ratio between the number of photons emitted to free space and the number of electrons injected into the LED) of 21.3% was obtained by Chiba et al[23] using red-emitting CsPb(Br/I)$_3$ and $EQE$=22% was recently reported by Dong et al[30] using CsPbBr$_3$ PNCs emitting in the green region ($EQE$=12.3% was obtained at 480 nm), in either case matching the record performance of analogous LEDs based on perovskite thin films[26,45]. Even in the best perovskite LED devices (PNC- or thin-film based) reported to date, the top limit to the $EQE$ was imposed by the extraction efficiency of the emitted photons from the device structure, $\eta_{OC}$, commonly referred to as light outcoupling, which is determined by the refractive index, absorbance, thickness and structure of the device layers[46-50].

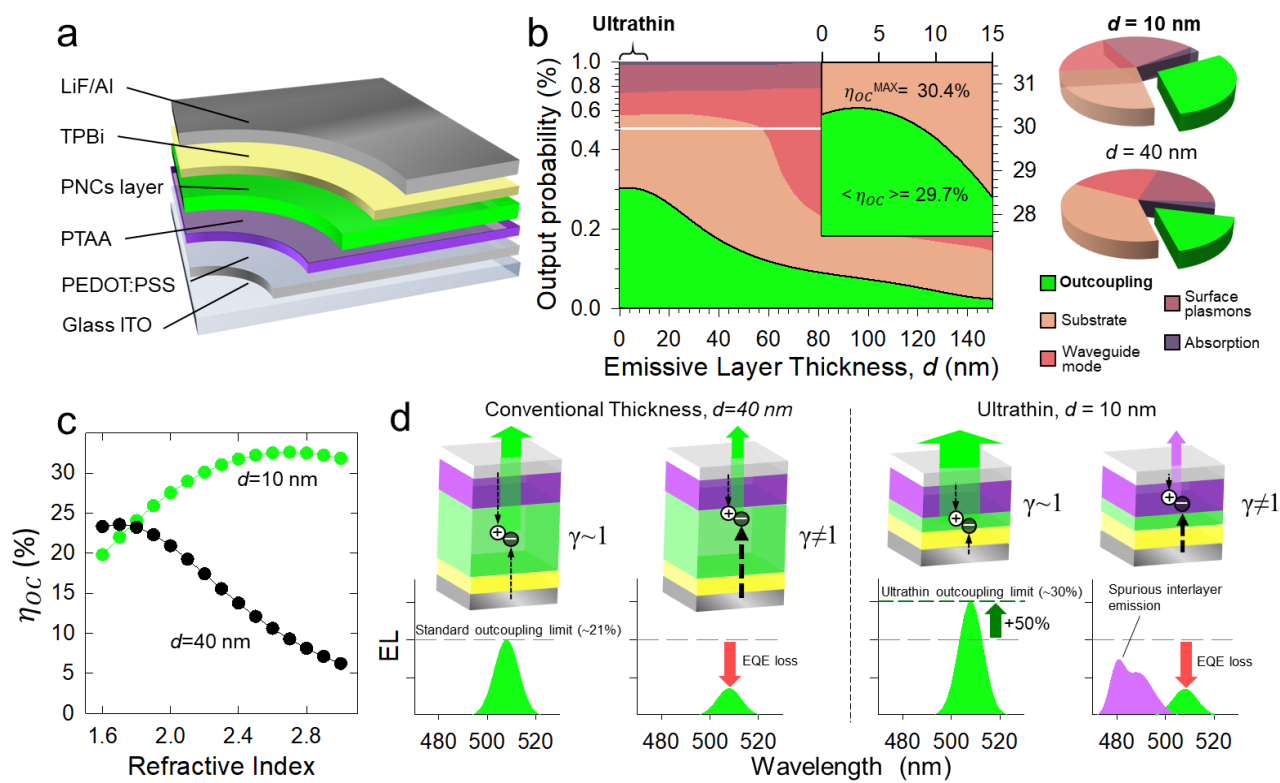

**Figure 1 a.** Schematic depiction of a PNC-LED with the architecture used for the light outcoupling simulation. **b.** Outcoupling probability distribution as a function the thickness of the emissive layer, $d$, for a PNC-LED with the same structure as in '**a**'. The shown output probability depends strongly on the refractive index and thickness of all functional layers, therefore particular care is suggested when comparing the simulated data with literature reports. **c.** $\eta_{OC}$ as a function of the refractive index of ultra-thin ($d$ =10 nm, green circle) and standard ($d$ =40 nm, black circles) active layer thickness. **d.** Illustration of the effect of the positioning of the carrier recombination zone in balanced vs. unbalanced LEDs with standard or ultra-thin active layers. For standard thickness devices, small charge imbalance ($\gamma\neq1$) lowers the efficiency, but the recombination zone is maintained within the PNC active layer thus preserving the color purity of the electroluminescence. In unbalanced ultra-thin LEDs, exciton formation and decay occur mostly in the charge regulating layer (the HTL in the depicted case) leading to severe parasitic emission (in purple) in addition to efficiency losses.

To highlight the dominant role of $\eta_{OC}$ in the $EQE$, it is instructive to theoretically simulate the maximum achievable performance (considering $IQE$=100%) of a planar device with the commonly employed architecture shown in **Figure 1a** as a function of the thickness of a CsPbBr$_3$ PNC emissive layer (indicated as $d$). In our calculations, we considered the same device architecture employed to realize the ultra-thin LEDs reported later in this work, featuring an indium tin oxide (ITO) coated (thickness 150 nm) glass anode coated with poly(3,4-ethylenedioxythiophene):polystyrene (PEDOT:PSS, 20 nm) / Poly(bis(4-phenyl)(2,4,6-trimethylphenyl)amine)

(PTAA, 20 nm) hole injection/ transport layers and 2,2′,2″-(1,3,5-benzinetriyl)-tris (1-phenyl-1-H-benzimidazole) (TPBi) (45 nm) electron transport layer covered with a LiF/Al (1 nm and 100 nm respectively) cathode. For the refractive index of the $CsPbBr_3$ PNC emissive layer, we employed n=2.18, as measured by ellipsometry (**Supplementary Figure S1**). In order to emphasize the effect solely of *d* without additional contributions from i.e. light scattering due to rough surface/interface morphology that artificially enhances light outcoupling, all layers were assumed as perfectly flat, also consistently with the very smooth film morphology of our real devices (*vide infra*). Consistent with previous reports, the simulation in **Figure 1b** shows that for *d* =35-60 nm, the maximum achievable *EQE* reaches ~21%, whereas the majority of the electroluminescence is lost due to the high refractive index of the perovskite material, parasitic contributions by the substrate, waveguiding modes, surface plasmons and light absorption of the interlayers.

For this reason, effective outcoupling strategies for the extraction of internal light are extremely powerful and highly sought-after for progressing PNC LED performances beyond the state-of-the-art. Several approaches have been successfully applied to organic or quantum-dot based LEDs[48,51-58], including the engineering of the refractive index, orientation of the emitter's dipole, the integration of photonic structures (i.e. micro-lenses arrays, diffraction gratings, low-index grids or buckling patterns inside the LED) or the optimization of the thickness of the charge-regulating layers so as to exploit microcavity effects. The use of patterned substrates or the spontaneous formation of submicron structures[14] have been found helpful for enhancing light outcoupling from organometallic perovskite LEDs[24,59,60]. Even though such motifs should, in principle, be helpful also for PNCs-LEDs, none of them has found successful integration to date.

An alternative approach to enhanced light extraction emerging from the theoretical simulation is thinning the emissive layer to *d*~10 nm (indicated here as the *ultra-thin regime*, **Figure 1b**). This should decrease the effective refractive index of the functional layers, substantially reducing the fraction of light trapped in waveguiding modes, ultimately enabling one to boost the *EQE* up to over ~30% without the need for complex photonic components (corresponding to ca. 50% enhancement with respect to standard thickness conditions). Importantly, as shown in **Figure 1c**, in the ultra-thin regime, $\eta_{OC}$ is found to slightly increase using emissive layers with high refractive index, in stark contrast to the substantial drop observed for thicker devices (i.e. *d* =40 nm vs. *d* =10 nm). This is an additional valuable benefit of this motif, as it opens to the possibility of exploring a wider range of materials with even higher refractive index than common halide perovskites (typically featuring $2.1 \leq n \leq 2.7$, depending on the halide composition, preparation method, etc.)[61,62], without sacrificing – and possibly further enhancing – the maximum achievable LED efficiency. Finally, the realization of ultra-thin PNC-LEDs would prompt their applicability to special situations where semi-transparency is a mandatory prerequisite, such as unobtrusive LED displays and screens, augmented reality, and wearable technology.

However, despite the great potential of the approach, the practical realization of highly efficient LEDs with ultra-thin PNC layers is extremely challenging, as it requires perfect equalization of the electron/hole currents inside the device, far beyond the operation conditions of common PNC-LEDs, and fine material processing to avoid the formation of pinholes in such a thin active layer. As sketched in **Figure 1d**, this is because small imbalances in charge injection/transport that are bearable for larger *d* values – although still lowering the EQE

–, in the ultra-thin regime have the additional damaging effect of locating the electron-hole recombination zone outside the emissive layer. This results in severe efficiency drop and in spurious parasitic electroluminescence from the organic charge-regulating layers, with consequent detriment of the LED color purity, as demonstrated by control experiments reported later in this work. Finally, being made up of a small number of PNCs, ultra-thin active films could be more affected by degradation (i.e. thermal, electrical etc.) than thick analogues, resulting in poor device stability. Therefore, in order to ripe the full potential of the ultra-thin regime to break the outcoupling limit, specific material design strategies are strictly needed.

In this work we aim to contribute to this challenge by practically realizing highly reproducible, stable, color-pure, and bright ultra-thin PNC-LEDs with *EQE* systematically exceeding the common outcoupling limit (average value <*EQE*>=24.6%, champion *EQE* as high as 29.2%) that, to the best of our knowledge, substantially overcome any perovskite-based LEDs reported to date. To achieve this regime, we took inspiration from the substantial improvement of the hole mobility and device stability observed in organic and quantum dot LEDs[63-66] and perovskite solar cells featuring nickel oxides hole injecting layers[67-72] to devise a post-synthesis $NiO_x$ resurfacing treatment that synergistically optimizes all three key parameters determining the *EQE* (namely $\gamma$, $\eta_{OC}$, and $\Phi_{PL}$). The application of this motif to both standard and advanced $CsPbBr_3$ PNCs, resulted in essentially perfectly equalized carrier injection and mobility that ultimately enabled us to engineer the recombination zone in active layers as thin as *d* =10 nm – nearly matching the low thickness limit set by the particle size – and to concomitantly enhance the emission yield and device operational stability.

**Results and Discussion**

For our study, we employed state of the art $CsPbBr_3$ PNCs synthesized using the recently reported triple ligand room temperature method[73], as well as 'control' PNCs of the same composition fabricated via conventional hot injection route, that further generalize the ultrathin strategy to overcome the common outcoupling limit. In both cases, the PNCs were resurfaced with a hole transporting $NiO_x$ layer using the multi-step post-synthesis treatment discussed in detail in the **Method Section**. Based on the nearly identical effect on the structural and physical properties of PNCs produced through the two different synthetic approaches, we focus hereafter on PNCs obtained via the triple ligand approach that yields the best device performance via a more convenient room temperature route, whereas the corresponding results on standard PNCs are reported in full in a dedicated section in the **Supplementary Material**. The complete study of the effects of surface functionalization on the structure and composition of either PNC sample, including transmission electron microscopy (TEM) images, size distribution, X-ray diffraction patterns and elemental mapping of pristine and $NiO_x$-treated $CsPbBr_3$ PNCs is reported in **Supplementary Figure S2-S4.** TEM images evidence no difference in PNC shape or size before and after resurfacing (average size of 10.4 ± 1.5 nm and 11.5±0.9 nm respectively) in both cases showing cubic lattice (with a possible slight shift to the orthorhombic phase in the treated particles) and the presence of nickel atoms adsorbed onto the particle surfaces, as confirmed by the elemental analysis and the surface etching experiment shown in **Figure S5**.

Fundamentally, the $NiO_x$ coating has noticeable beneficial effects on the electrical and optical properties of the PNCs, which are both crucial for the realization of highly efficient ultra-thin LEDs. Specifically, as shown by the flat-band energy diagram of pristine and CsPbBr$_3$:NiO$_x$ PNCs (**Figure 2a**) estimated from ultraviolet photoelectron spectroscopy (UPS) measurements (**Supplementary Figure S6**), the NiO$_x$ treatment produces a rigid raise (~400 meV) of both frontier levels of CsPbBr$_3$ PNCs. This substantially lowers the energy offset between the HOMO level of PTAA ($E_{HOMO}$=-5.25 eV) and the valence band of the PNCs ($E_{VB}$=-5.4 eV), thus facilitating hole injection – the effect on electron injection is negligible due to the already favorable potential offset between the LUMO of TPBi and the PNCs conduction band. Importantly and in agreement with previous studies on organic and quantum dot LEDs using nickel oxides interlayers to enhance hole injection and transport[63-65], carrier mobility measurements by space-charge limited current (SCLC) method in single carrier devices (**Figure 2b**) show that the hole mobility increases from the typical value[23,74] of $\mu_h$=0.98±0.10×10$^{-3}$cm$^2$V$^{-1}$s$^{-1}$ to $\mu_h$=1.51±0.18×10$^{-3}$cm$^2$V$^{-1}$s$^{-1}$ upon NiO$_x$ resurfacing, closely matching the respective electron mobility, which also grows from $\mu_e$=1.26±0.18×10$^{-3}$cm$^2$V$^{-1}$s$^{-1}$ to $\mu_e$=1.52±0.16×10$^{-3}$ cm$^2$V$^{-1}$s$^{-1}$. Altogether, this results in essentially perfect carrier mobility balance in devices based on CsPbBr$_3$:NiO$_x$ PNCs, as quantified by the term $\gamma=\mu_e/\mu_h$ that approaches unity ($\gamma$=1.08±0.07) with respect to the pristine case featuring $\gamma$=1.29±0.06. Due of the initially lower mobilities in conventional PNC solids, the mobility balance increases even more substantially for the standard CsPbBr$_3$ PNCs after NiO$_x$ treatment, for which $\gamma$ changed from 1.50±0.04 in the pristine case to $\gamma$ =1.09±0.08.

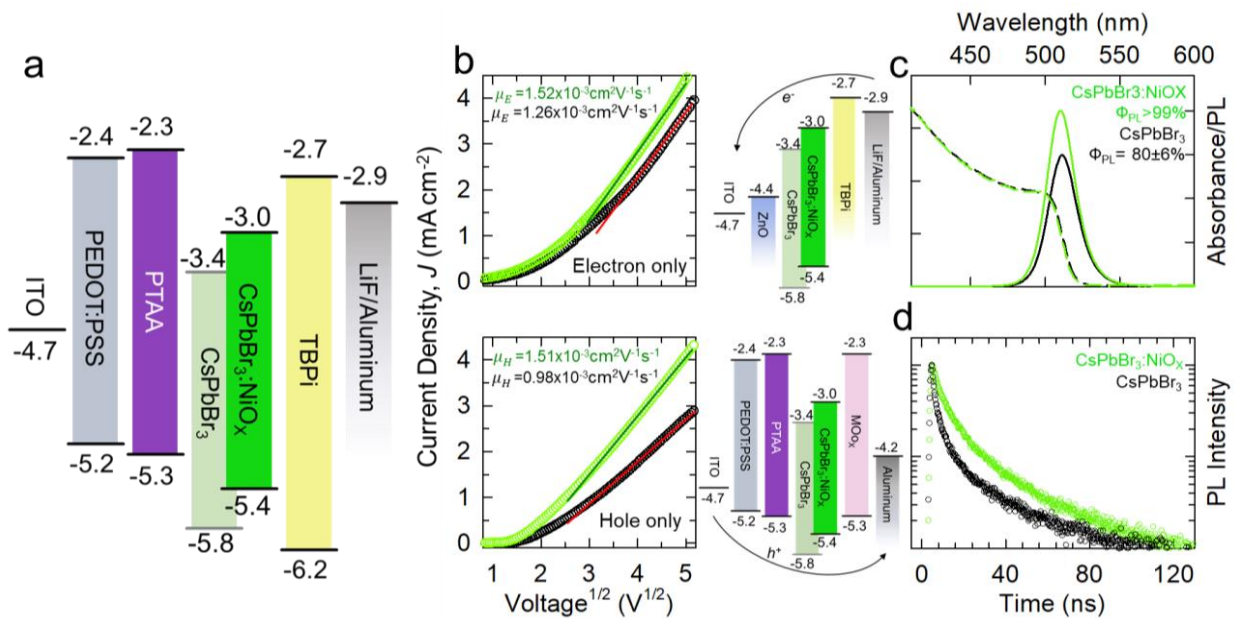

**Figure 2 a.** Flat band energy diagram of a multilayer LED highlighting the shift of the frontiers level of CsPbBr$_3$ PNCs upon treatment with NiO$_x$. **b.** Carrier mobility (top panel: electrons, lower panel: holes) in single carrier devices embedding pristine CsPbBr$_3$ PNCs (black symbols) or CsPbBr$_3$:NiO$_x$ PNCs (green symbols). **c.** Optical absorption (dashed lines) and photoluminescence (solid lines) spectra of pristine (black) and NiO$_x$-treated CsPbBr$_3$ PNCs (green) in toluene solution (excitation at 450 nm) and, **d.** respective photoluminescence decay curves.

We notice that this effect is also likely corroborated by the PNC washing process and by the effective passivation of surface carrier traps posited by the NiO$_x$ coating. Suppression of surface trapping is confirmed by the side-by-side spectroscopic analysis shown in **Figure 2c** and **2d**. Upon NiO$_x$ resurfacing, the photoluminescence quantum yield grows from $\Phi_{PL}$=80±6% of the pristine PNCs to $\Phi_{PL}$>99% ($\Phi_{PL}$=48±6% to 95±6% for the standard PNCs), accompanied by a noticeable lengthening of the decay kinetics. Also in agreement with the rigid shift of both energy levels in CsPbBr$_3$:NiO$_x$ PNCs (**Figure 2a**), the bandgap energy of the PNCs is maintained upon treatment, thus leaving the spectral position of the first excitonic absorption feature and luminescence peak essentially unaltered.

We then proceeded with experimentally validating the potential of the ultra-thin emissive layer approach by fabricating and testing planar devices with the architecture shown in **Figure 1a** embedding an ultrathin PNC emissive layer ($d$ = 10 nm). The cross sectional TEM image is reported in **Figure 3a**, showing a continuous film of CsPbBr$_3$:NiO$_x$ PNCs with clear interfaces. The surface morphology of the interlayers and of the ultra-thin PNC film deposited onto ITO/PEDOT:PSS/PTAA was investigated by atomic force microscopy revealing uniform, continuous films with RMS roughness as low as 0.63 nm and 2.04 nm respectively (**Supplementary Figure S7**). As anticipated above, the smoothness of the interfaces is important for our discussion as it minimizes light scattering effects on light outcoupling, consistent with the simulations in **Figure 1b**. The whole device shows an ultralow thickness of about 400 nm (excluding the glass substrate) resulting in substantial transparency across the visible spectrum, as shown by the photograph of the complete device, except for the metal cathode, reported in the inset of **Figure 3a** and by the corresponding transmission spectrum in the inset of **Figure 3b**. Consistently, the color rendering index of transmitted light from D65 standard illuminant[75] is as high as R$_a$=95. Although the engineering of fully transparent LEDs including the cathode material is beyond the scope of this study, **Supplementary Video 1** shows an example of a functioning ultra-thin LED with a semitransparent cathode (20 nm) emitting light from both sides. In **Figure 3b** and **3c** we report the current density–voltage ($J$-V) and luminance-voltage ($L$-V) responses of a representative LED embedding an ultrathin layer of CsPbBr$_3$:NiO$_x$ PNCs ($d$ =10 nm). The LED shows low turn on voltage ($V_{ON}$, defined as the voltage at which the luminance is 1 cd/m$^2$) of 2.8 V, high luminance up to ~10$^4$ cd/m$^2$ and remarkable EQE= 26.5% which substantially exceeds the outcoupling limit of conventional devices. External confirmation of the result was provided by independent measurements on the same device, also certifying the very low efficiency roll-off despite the low thickness of the emissive layer. The repeatability of our result over 40 identical devices (featuring $d$ = 10 nm) is highlighted in **Figure 3d** showing that all produced LEDs exceeded the outcoupling limit for standard PNC-LEDs and thus systematically outcompeted any perovskite-based LEDs with emission in any spectral region reported to date, with average <$EQE$>=24.6±1.9% and champion $EQE_C$= 29.2%, corresponding to nearly 40% increase with respect to the maximum efficiency in the conventional outcoupling limited regime (~21%). The statistics over the same number of devices produced using conventional CsPbBr$_3$ PNCs resurfaced with NiO$_x$ is also reported showing, also in this case, very high EQE systematically exceeding the outcoupling limit ($EQE_C$= 27.3%, <$EQE$>=23.9±1.8%), yet with lower average value due to the better quality of PNCs produced using the triple ligand approach. Crucially, at any driving voltage (up to 7V), the electroluminescence spectrum features exclusively a narrow (FWHM=20 nm) peak at 514 nm (**Figure 3e**), matching the respective PL profile except for a minor, 4 nm, red shift due to Stark effect.

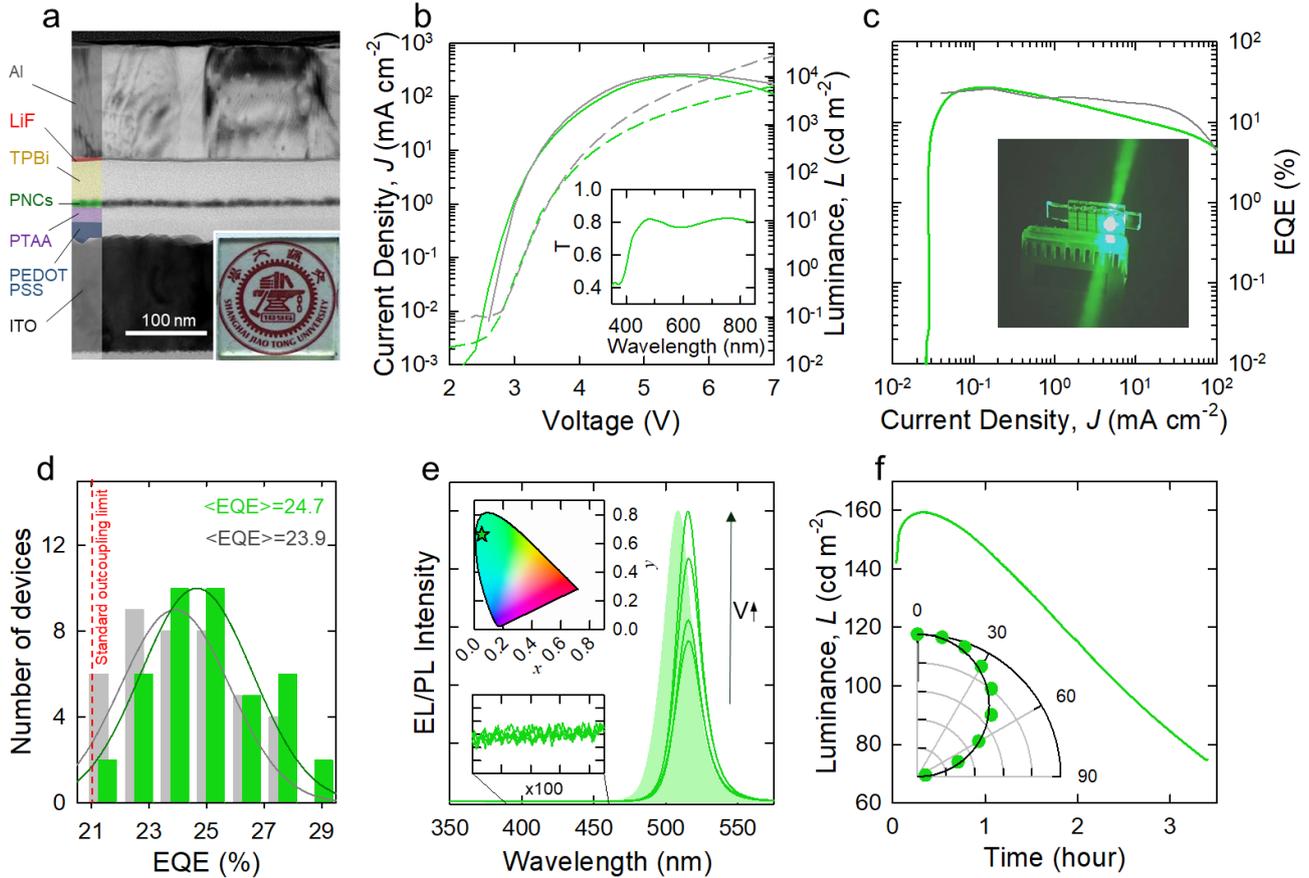

**Figure 3 a.** Cross sectional TEM images of a LED sample embedding an ultra-thin active layer of $CsPbBr_3:NiO_x$ PNCs. **b.** Current density – Voltage (*J*-V, dashed lines) and Voltage – Luminance (L-V, continuous lines) curves of a representative $CsPbBr_3:NiO_x$ PNCs-based LED measured in our lab (green lines) and independently at the Research Center of Photoelectric Materials and Devices by the Department of Electronic Engineering of the Shanghai Jiao Tong University (grey lines). The characterization was performed at room temperature in air with simple device encapsulation. Inset: Transmission spectrum of the LED (*d* = 10 nm) without the metal cathode. **c.** *EQE* curves of the same LED reported in 'b' (same color code as in b). Inset: photography of a $CsPbBr_3:NiO_x$ PNCs-based LED (active area 4 mm$^2$). **d**. *EQE* statistics over 40 identical ultra-thin LEDs (*d* = 10 nm) embedding $CsPbBr_3:NiO_x$ PNCs synthetized through room temperature approach (green bars) and hot injection method (grey bars). **e.** EL spectra (dark green curves) of an ultra-thin LED embedding $CsPbBr_3:NiO_x$ (*d* = 10 nm) at increasing bias voltage (3-7 V as indicated by the arrow) together with the respective PL spectrum (shaded curve, excitation wavelength 450 nm). Bottom Inset: enlargement (×100) of the 380-470 nm region showing no emission by the organic interlayers. Top Inset: Chromaticity coordinates corresponding to the EL spectra of the ultra-thin PNC-LED based on $CsPbBr_3:NiO_x$ shown in the CIE chromaticity diagram, highlighting the vicinity of the color coordinates to the spectral locus. **f.** Luminance as a function of operating time of a representative $CsPbBr_3:NiO_x$ PNCs-based LED tested at room temperature in air with simple encapsulation. Inset: Angular emission profile of the same LED. The black line indicates the expected angular intensity distribution for a perfectly Lambertian emitter.

The corresponding Commission Internationale de l'Eclairage coordinates are *x*=0.05 and *y*=0.66, corresponding to ~90% saturated green light. No spurious emission from the organic interlayers is observed, which confirms that the recombination zone is fully confined in the ultra-thin active layer and that the emissive layer is a continuous film of PNCs even at such a low thickness level. Devices with gradually increasing *d*, showed progressively higher $V_{ON}$ from 2.8 eV to 5.2 V accompanied by the drop of the luminance and EQE below the respective maximum theoretical value, which indicates that the *IQE* decreases in thicker NCs films, likely due to electrical losses in thicker PNC layers, which are not taken into account in the optical simulation (**Supplementary Table S1**).

More importantly, despite their very low thickness, our LEDs with ultrathin active layer exhibited operational luminance stability comparable to the best devices reported to date featuring much thicker emissive films, with half-lifetime ($T_{50}$, defined as the time after which the luminance decreases of a factor of two) as long as $T_{50}$=3 hours when operated at constant current density of 1 mA/cm$^2$ (corresponding to initial luminance $L_0$~150 cd m$^{-2}$, **Figure 3f**). Such a remarkable stability is likely due to the combined beneficial effect of the triple ligand approach used for the PNC synthesis and to the NiO$_x$ resurfacing, as supported by the substantially lower stability of comparable LEDs embedding pristine CsPbBr$_3$ NCs (**Supplementary Figure S8**) and in agreement with recent reports on highly stable perovskite solar cells with nickel oxide interlayers[68,72]. Finally, angular resolved electroluminescence measurements of our ultrathin LED featuring a CsPbBr$_3$:NiO$_x$ PNCs active layer showed Lambertian emission distribution (inset of **Figure 3f**), thus indicating that exceeding the outcoupling limit is not due to microcavity effects[76] in agreement with our theoretical prediction. We emphasize that ultra-thin LEDs with the performance shown in **Figure 3** cannot be achieved with simple pristine CsPbBr$_3$ PNCs produced with either synthetic approach. As shown in **Supplementary Figure S9**, already at conventional thicknesses ($d$ =40 nm), control LEDs based on pristine PNCs exhibit only half the EQE and maximum luminance than analogues devices featuring NiO$_x$ treated particles. Furthermore, because of disproportionate charge injection and transport and of discontinuities in the film structure, the electroluminescence spectrum of a control LED with $d$ < 12 nm features a strong blue emission emerging from carrier recombination in the hole transport layer **Supplementary Figure S10**. The whole characterization of the control devices is reported in **Supplementary Figure S11** and **Supplementary Table S2**, showing, in agreement with previous results[12,32], that the thinning approach also helps improving the efficiency of conventional PNC-LEDs that, however, feature overall lower performance due to limited optical and electrical properties of standard CsPbBr$_3$ PNCs.

In conclusion, we have demonstrated that thinning the active layer to 10 nm is an effective strategy to systematically enhance the efficiency of PNC-LEDs above the maximum value imposed by light outcoupling in standard devices, reaching *EQE* values exceeding 29%. The approach, demonstrated here using CsPbBr$_3$ PNCs produced via both room temperature and hot injection methods, is general and, in principle, applicable to other perovskite nanostructures for fabricating highly efficient, color tunable LEDs. It is also compatible with the artificial introduction of outcoupling photonic structures for further boosting the device performance. These results offer a potentially powerful strategy for high efficiency perovskite LEDs and a valuable route for next generation display and lighting technologies.

**Author information**
Qun Wan, Weilin Zheng and Chen Zou contributed equally to this work.

**Corresponding Author**
*E-mail: liangli117@sjtu.edu.cn, sergio.brovelli@unimib.it

**Acknowledgments**
This work was supported by Guangdong Province's Key R&D Program (2019B010924001), Shanghai Jiao Tong University Scientific and Technological Innovation Founds, and the National Key R&D Program of China (No.2018YFC1800600). We gratefully acknowledge financial support from the Italian Ministry of University and Research (MIUR) through the grant Dipartimenti di Eccellenza-2017 'Materials for Energy'. Here, we want to give a special thanks to Professor Gufeng He in the Research Center of Photoelectric Materials and Devices for his help to confirm the performance of device, and Professor Baoquan Sun and Dr. Yatao Zou in Soochow University for their help to measure the angular distribution of emission pattern. We also thank the instrumental analysis center of Shanghai Jiao Tong University and the workers (Zhongqiu Bao, Ying Zhang, Yao Han, Chong Lu, Xinqiu Guo and Jiaxin Ding) for the help in SEM, TEM and TOF SIMS measurements. Center for Advanced Electronic Materials and Deices of Shanghai Jiao Tong University also supported for the measurements of ellipsometry spectrum (Semilab SE-2000) and film thickness (KLA-Tencor P7).

**Competing interests**
The authors declare no competing financial interests.

**Additional information**
Supplementary information is available for this paper at

**Reprints and permissions information is available at www.nature.com/reprints.**

**Correspondence and requests for materials** should be addressed to L.L. and S.B. The data that support the plots within this paper and other finding of this study are available from the corresponding author upon reasonable request.

**Publisher's note:** Springer Nature remains neutral with regard to jurisdictional claims in published maps and institutional affiliations.

# Supplementary Information

# Highly Efficient Ultrathin Light Emitting Diodes based on Perovskite Nanocrystals

Qun Wan[a,§], Weilin Zheng[a,§], Chen Zou[b,§], Francesco Carulli[c] Congyang Zhang[b], Haili Song[d], Mingming Liu[a], Qinggang Zhang[a], Lih Y. Lin[b], Long Kong[a], Liang Li[a*] and Sergio Brovelli[c*]

[a]*School of Environmental Science and Engineering, Shanghai Jiao Tong University, 800 Dongchuan Road, 200240 Shanghai, P.R. China*

[b]*Department of Electrical and Computer Engineering, University of Washington, Seattle, WA 98195, U.S.A*

[c]*Dipartimento di Scienza dei Materiali, Università degli Studi di Milano-Bicocca, 20125 Milano, Italy*

[d]*Key Laboratory of Bioinorganic and Synthetic Chemistry of Ministry of Education, School of Chemistry, Sun Yat-Sen University, Guangzhou 510275, China*

**Materials and Methods**

*Materials.* Cesium carbonate ($Cs_2CO_3$, 98%, Macklin), 1-octadecene (ODE, 90% Aladdin), oleic acid (OA, 90% Aladdin), octanoic acid (OTAc, 99%, Macklin), oleylamine (OAm, 90% Aladdin), lead bromide ($PbBr_2$, 98%, Aladdin), didodecyl dimethylammonium bromide (DDAB, 98%, Aladdin), tetraoctylammonium bromide (TOAB, 98%, Macklin), methyl acetate (98%, Aladdin), toluene (98%, Aladdin), nickel acetate terahydrate (Ni(Ac)$_2$·4H$_2$O, 99.9%, Aladdin), benzoyl peroxide (BPO, 98 %, Aladdin) and 1-octane (99%, Sigma-Aldrich) were used without further purification. Poly(3,4-ethylenedioxythiophene):polystyrene (PEDOT:PSS), 4-Butyl-N,N-diphenylaniline (Poly-TPD) and 2,2',2''-(1,3,5-Benzinetriyl)- tris(1-phenyl-1-H-benzimidazole) (TPBi) were purchased from Xi'an Polymer Light Technology Corp. Poly(bis(4-phenyl)(2,4,6-trimethylphenyl)amine) (PTAA) was purchased from LinkZill Technology Crop.

*Preparation of CsOA precursor:* A 100 mL round-bottom flask equipped with a reflux condenser and a thermocouple probe was loaded with 10 mmol of $Cs_2CO_3$, 20 mL of ODE, and 20 mL of OA under standard air free conditions. The reaction system was evacuated for 1h at 120 °C, and then the temperature was raised to 150 °C under argon flow to form an optically clear solution. After 0.5 h, the solution was cooled down to room temperature, and stored as stock solution.

*Synthesis and Purification of $CsPbBr_3$ PNCs prepared via room temperature triple ligand method.* The $CsPbBr_3$ PNCs were synthesized following an adaptation of the literature recipe. Briefly, 1 mL $Cs_2CO_3$ solution (0.1 mmol mL$^{-1}$ in OTAc) was quickly added into 9 mL $PbBr_2$ solution (0.1 mmol mL$^{-1}$ in toluene, mixed with TOAB as a molar ratio of 1:2). The solution was stirred for 5 min at room temperature in air. Then, 3 mL DDAB solution (10 mg mL$^{-1}$ in toluene) was added into the solution and stirred for another 2 min. The $CsPbBr_3$ PNCs solution was obtained through washing, centrifugation and re-dispersion. The $CsPbBr_3$ PNCs used for device fabrication were re-dispersed in n-octane.

*Synthesis of control $CsPbBr_3$ PNCs prepared via hot injection route:* 20mL of ODE, 5 mL of OAm, 5 mL of OA, and 2 mmol $PbBr_2$ were loaded into a 100 mL three-neck flask, degassed at 120 °C for 30 min. Then, the temperature was raised to 180 °C under argon atmosphere until complete dissolution of $PbBr_2$ salt. Afterwards, 1 mL of CsOA precursor (0.5 M), pre-heated at 70°C, was injected into the flask. After 10 s, the three-neck flask was placed in an ice-water bath and cooled to room temperature. The crude solution was precipitated by methyl acetate and toluene via centrifugation. The precipitate was finally dispersed in a n-octane solution for processing.

*DDAB treatment on pristine $CsPbBr_3$ PNCs:* 1 mL DDAB toluene solution (0.02 mmol/mL) was added into 10 mL of the purified $CsPbBr_3$ PNCs (ca. 15 mg/mL), then the mixture solution was stirred for 2 h under room temperature. After treatment, the PNCs were purified with methyl acetate, and the $CsPbBr_3$-DDAB PNCs were obtained.

*Preparation of $CsPbBr_3$:$NiO_X$ PNCs:* The $NiO_X$ resurfacing of PNCs was performed through a two-step process: initially Ni(Ac)$_2$ (6.5 mg) was added to a 10 mL $CsPbBr_3$-DDAB PNCs solution (molar ratio of Ni(Ac)$_2$ to DDAB was 1.25:1). The solution was stirred for 2 hours and the resurfacing procedure was completed with the slow addition of benzoyl peroxide (BPO) in toluene solution (0.05 mmol/mL) into the $CsPbBr_3$-DDAB:Ni PNCs solution at 60 °C within 2 h. After treatment, the NCs were purified with methyl acetate, and the $CsPbBr_3$:$NiO_X$ PNCs were washing, centrifugation and re-dispersed in n-octane.

*Device Fabrication:* The devices were fabricated with the following structure: indium oxide (ITO)/ PEDOT:PSS (~35 nm or 20 nm) / Poly-TPD (~30 nm) or PTAA (~20 nm) / PNCs / (TPBi) (~40 nm or 45 nm) /LiF (~1 nm) /Al (~100 nm). PEDOT:PSS solution were spin-coated onto the ITO-coated glass substrates and baked at 145°C for 15min. The substrates were then transferred into a N$_2$-filled glovebox. The hole transporting layer was prepared by spin-coating Poly-TPD (concentration of 8 mg/mL) or PTAA (concentration of 5 mg/mL) from a chlorobenzene solution. PNCs layers were deposited by spin coating of solutions with different concentrations and baking at 60 °C for 10 min. TPBi and LiF/Al electrodes were deposited using a thermal evaporation system through a shadow mask under a high vacuum. The device active area was 0.04cm$^2$ as defined by the overlapping area of the ITO and Al electrodes.

*Electron-only device fabrication:* Cleaned ITO substrates were treated by UV-ozone for 15 min. A solution of ZnO particles (~30 mg mL$^{-1}$) was spin-coated on the ITO glass at 2000 rpm for 20s and annealed at 90°C for 30 min. Then PNCs layer was spin coated. TPBi (40 nm), LiF (1 nm) and Al (100 nm) were sequentially deposited by thermal evaporation in a vacuum deposition chamber (~2×10$^{-4}$ Pa).

*Hole-only device fabrication:* Cleaned ITO substrates were treated by UV-ozone for 15 min. PEDOT:PSS solution was spin-coated onto the ITO glass at 4000 rpm for 20 s and annealed in air at 145°C for 15 min. Then the substrate was transferred into the glovebox. Poly-TPD or PTAA layer was prepared by spin-coating the solution and annealed at 110 °C or 120 °C for 20 min. PNCs layer was spin cast from n-octane dispersions at 2000 rpm. MoO$_3$ (10 nm) and Al (100 nm) were the sequentially deposited by thermal evaporation in a vacuum deposition chamber (~2×10$^{-4}$ Pa).

*Morphological and structural Characterizations:* The morphology and elemental mappings of the PNCs were observed with a FEI (TALOS F200X) transmission electron microscope (TEM) instruments. The valence band (VB) spectra were measured with a monochromatic He I light source (21.2 eV) and a VG Scienta R4000 analyzer.

*Optical Characterization:* The optical absorption spectra of PNC solutions were measured using a Hitachi U-3900 spectrophotometer. The PL spectra were measured through a Hitachi F-7000 fluorescence spectrophotometer. The EL spectra of the PNCs-LEDs were measured using an Ocean Optics USB 2000 spectrometer. Time-resolved PL measurements were carried out by Edinburgh Instruments FLS1000 spectrometer with a 365 nm excitation light source. The absolute PL quantum yields of the PNCs solutions were measured by exciting at 395 nm using a LED light source and collecting the fluorescence with a CCD spectrometer coupled with an integrating sphere.

*LED Characterization:* The current density-voltage-luminance (J-V-L) characteristics were measured simultaneously with a Keithley 2401 Sourcemeter® (current measurement accuracy of 10pA) and Konica Minolta LS-160 luminance meter in air and at room temperature. The device edges were carefully blocked with black paint so as to avoid collecting guided light.

*Single-carrier conductivity characterizations:* The single carrier device structures of electron and hole were ITO/ZnO/PNCs/TPBi/LiF/Al and ITO/PEDOT:PSS/Poly-TPD or PTAA/PNCs/MoO$_3$/Al, respectively. The mobility of electron and hole were calculated as the following law

$$J = \frac{9}{8}\varepsilon_r\varepsilon_0\mu\frac{V^2}{d^3}.$$

Where J is the current density, $\varepsilon_r$ and $\varepsilon_0$ are the relative dielectric constant and vacuum dielectric constant, respectively. And $\mu$ is the field-independent carrier mobility, and d is the thickness of the PNCs layer. Here, the values of $\varepsilon_r$, and $\varepsilon_0$ are 28 and 8.85×10$^{-14}$ F/cm respectively, while and *d* was considered 30 nm.

*Outcoupling Simulation:*

The light outcoupling of PNCs-based LEDs were simulated using a transfer matrix method (TMM) and classical dipole mode [10.1364/OL.400814] for a device with same structure reported. In this model, in accordance with the experimental measurements, we assumed the intrinsic quantum yield of PNCs thin film is 100% and the charge injection is well balance with γ = 1. The refractive index of the CsPbBr$_3$ film is ≈ 2.2 near the emission wavelength. The emission zone is assumed to be infinitely thin and located at the middle of the perovskite layer. The emitting medium is assumed nonabsorbing at the emissive region in the simulation for numerical stability [Nature Photonics 15, 148-155 (2021), Advanced Optical Materials 6, 17 (2018): 1800667].

In general, the optical power generated in PNCs-based LED is distributed to 4 channels, separated by the in-plane wavevector $k_{//}$.

(1) Direct emission: $k_0 \cdot n_{air} \geq k_{//} \geq 0$, where $k_0 = 2\pi/\lambda$ is the vacuum wave vector and $n_{air}$ is the refractive index of air. In this region, light outcouples into the air from PNCs.

(2) Substrate mode: $k_0 \cdot n_{sub} \geq k_{//} \geq k_0 \cdot n_{air}$, where $n_{sub}$ is the refractive index of the substrate. In this region, light is trapped in the substrate due to the total internal reflection (TIR) at the substrate and air interface.

(3) Waveguide mode: $k_0 \cdot n_{eff} \geq k_{//} \geq k_0 \cdot n_{sub}$, where $n_{eff}$ is the effective refractive index of the functional layers. In this region, light is trapped in functional layers because of the TIR at the ITO and substrate interface.

(4) Surface plasmon mode: $k_{//} \geq k_0 \cdot n_{eff}$. In this region, light couples to the top metal electrode in the form of evanescent waves.

(5) Absorption: for the four optical channels, part of the dissipated power is absorbed inside the device instead of leaking into corresponding optical channels.

| Layer | Thickness | Refractive index (n) |
|---|---|---|
| Al/LiF | 100nm/1nm | 0.8+i*6.2 |
| TPBi | 45 nm | 1.73 |
| PNCs | 1-300 nm | 2.2 |
| PTAA | 20 nm | 1.74 |
| PEDOT:PSS | 20 nm | 1.51+i*0.02 |
| Glass-ITO | 150 nm | 1.87+i*0.007 |

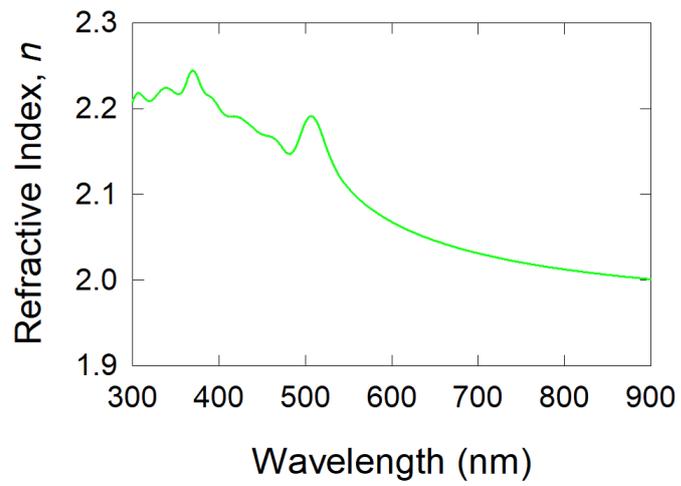

**Figure S1**: Ellipsometry measurement of a $CsPbBr_3$:$NiO_X$ PNCs sample synthetized though the triple ligand room temperature route.

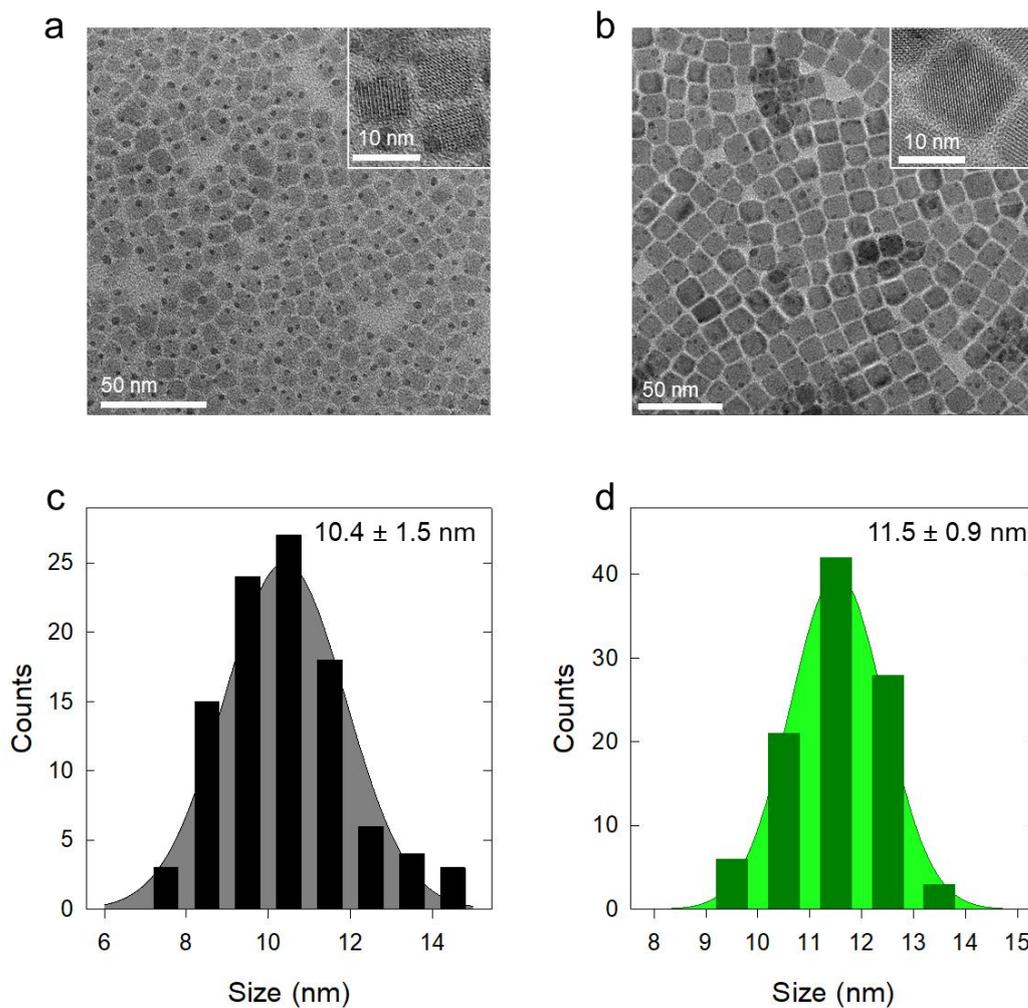

**Figure S2**. TEM images of a) pristine $CsPbBr_3$ PNCs and b) $CsPbBr_3$:$NiO_X$ PNCs prepared via room temperature triple ligand procedure. Size distributions of c) pristine $CsPbBr_3$ PNCs and d) $CsPbBr_3$: $NiO_X$ PNCs. Average size and standard deviation of both samples are reported.

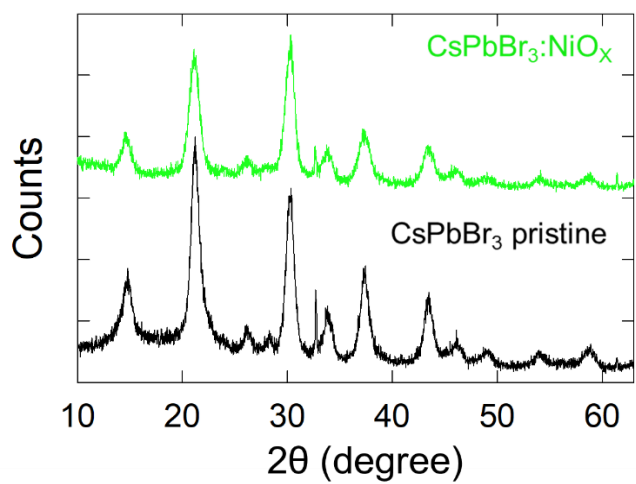

**Figure S3**: X-Ray diffraction patterns of pristine CsPbBr$_3$ PNCs (black curve) and CsPbBr$_3$:NiO$_X$ PNCs (green curve) prepared via room temperature triple ligand procedure.

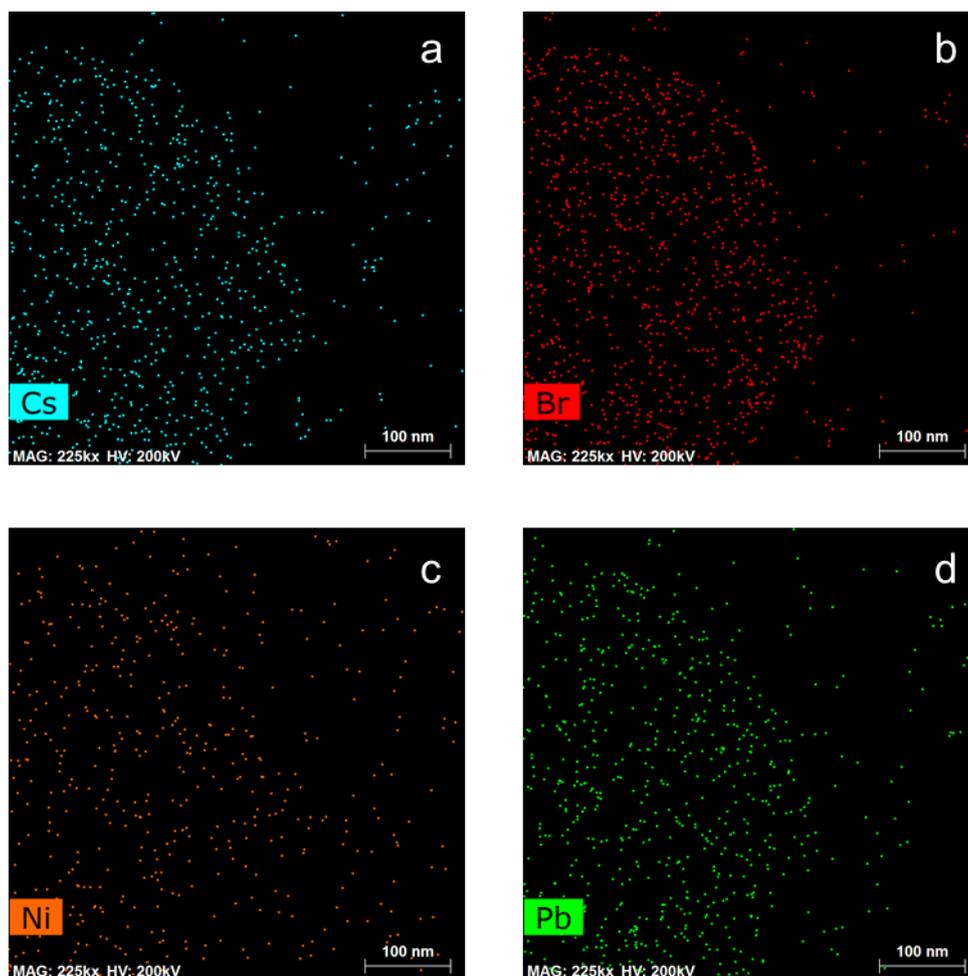

**Figure S4.** Elemental mapping images of Cesium (a), Bromine (b), Nickel (c) and Lead (d) in a CsPbBr$_3$:NiO$_X$ PNCs sample prepared via room temperature triple ligand procedure.

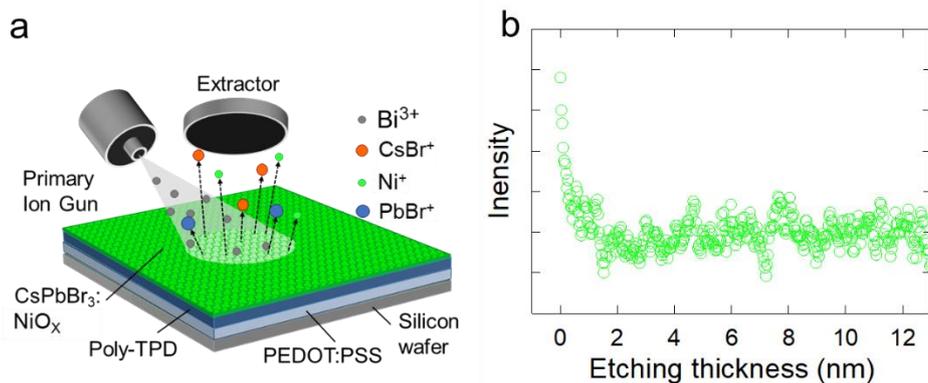

**Figure S5.** Schematic diagram of TOF SIMS set up (a) and intensity of Ni$^+$ ions related signal as a function of the etching depth on a CsPbBr$_3$:NiO$_X$ PNC sample.

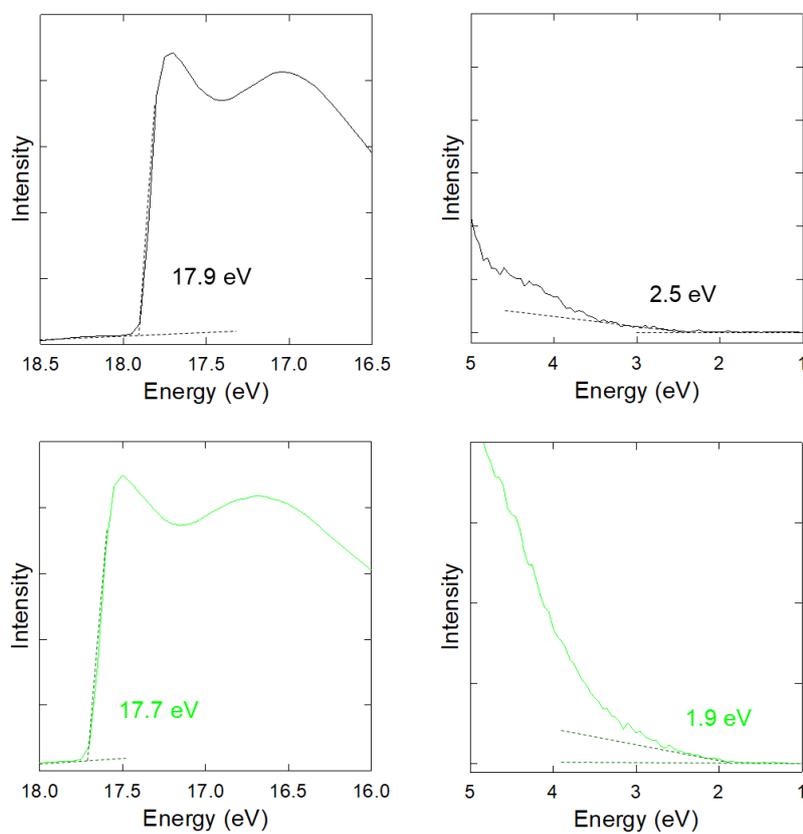

**Figure S6**. UPS spectra of pristine CsPbBr$_3$ (black line, top plots) and CsPbBr$_3$:NiO$_X$ (green line, bottom plots) samples prepared via the room temperature triple ligand route.

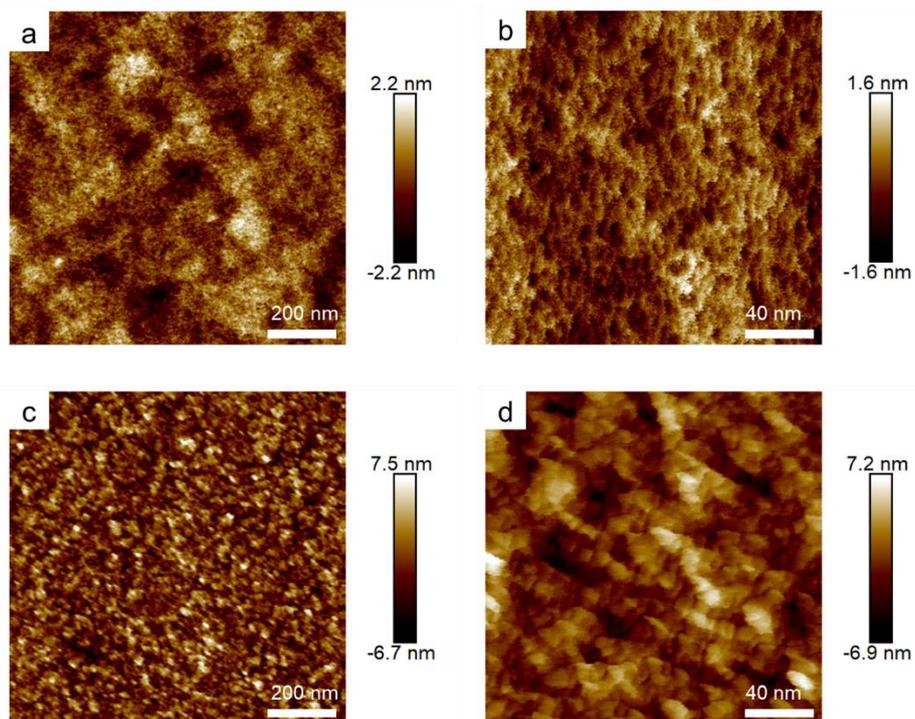

**Figure S7**. AFM images of ITO/PEDOT:PSS/PTAA (top panels) and ITO/PEDOT:PSS/PTAA/CsPbBr$_3$:NiO$_X$ (bottom panels) measured in height mode.

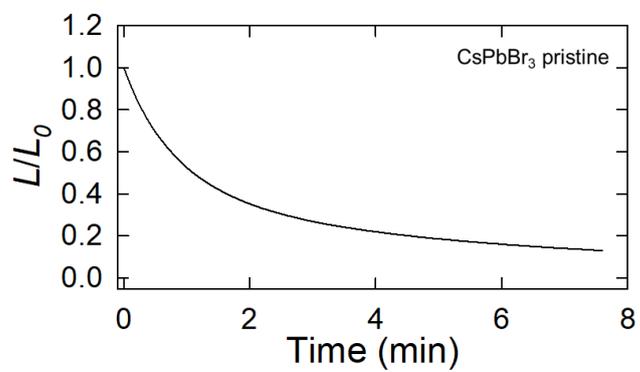

**Figure S8**. Luminance as a function of operating time of a representative pristine CsPbBr$_3$ PNCs-based LED.

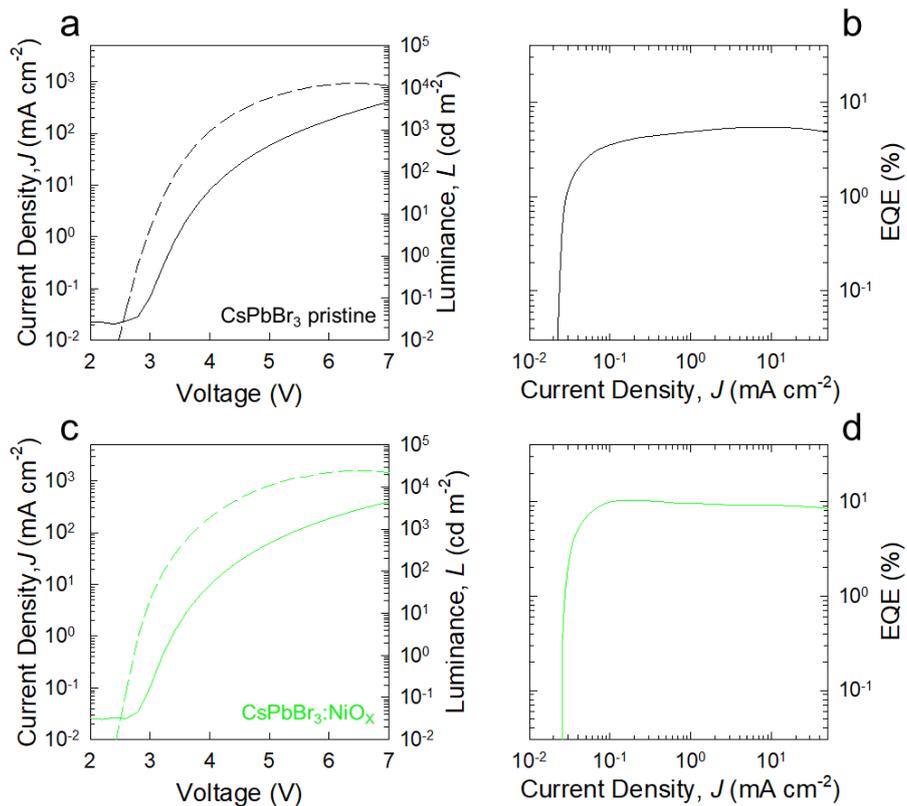

**Figure S9.** Current density – Voltage (J-V, continuous line), Voltage – Luminance (L-V, dashed line) (a and c plots) and EQE curves (b and d plots) of LEDs with $d$ =40 nm thick emissive layers of pristine $CsPbBr_3$ and $CsPbBr_3$:$NiO_X$ PNCs respectively.

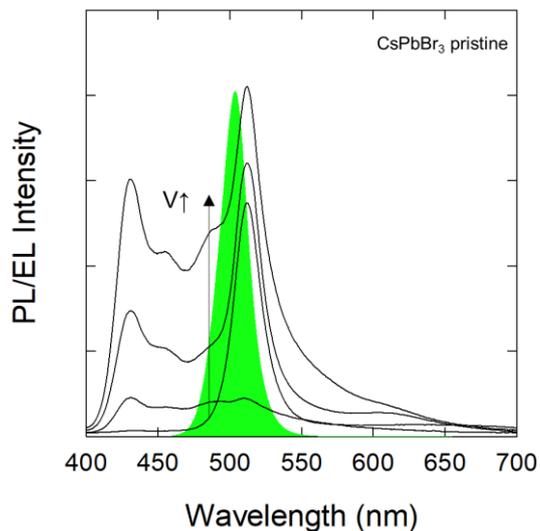

**Figure S10.** EL spectra of a pristine $CsPbBr_3$ PNCs-based device at increasing voltage (4-7 V). PL spectra of pure $CsPbBr_3$ PNCs is reported (green shaded area) as comparison.

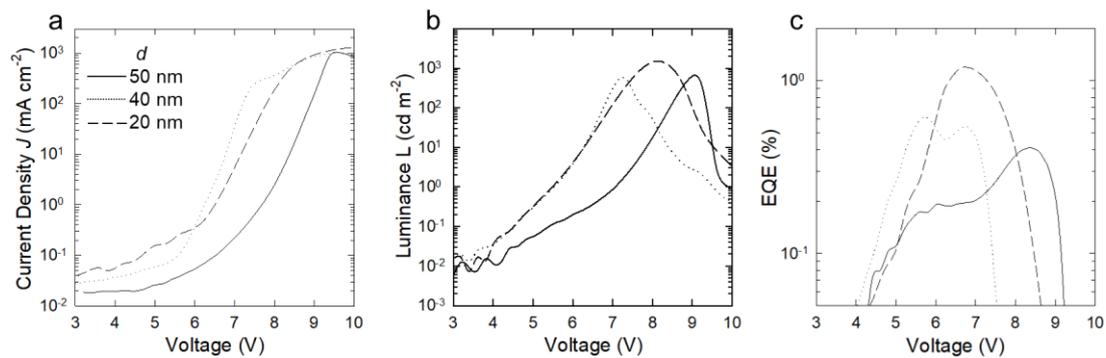

**Figure S11.** a) Current density-voltage (*J*-V), b) Luminance-voltage (L-V) and EQE-current density (EQE-*J*) characteristics of LEDs based on pristine CsPbBr$_3$ PNCs with different active layer thickness. The same line code applied to all plots.

**Characterization of PNCs and PNCs-based LEDs synthetized through a standard hot injection route**

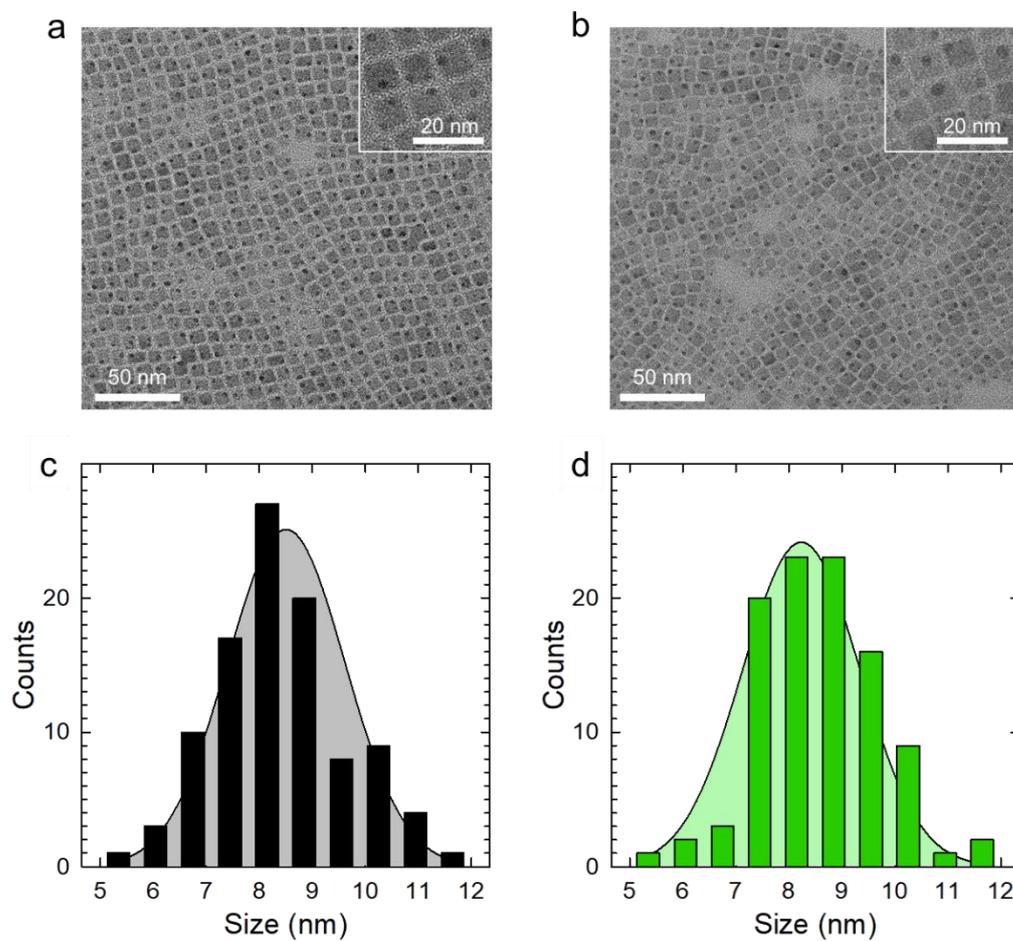

**Figure S13**. TEM images of a) pristine $CsPbBr_3$ PNCs and b) $CsPbBr_3$:$NiO_X$ PNCs. Size distributions of c) pristine $CsPbBr_3$ PNCs and d) $CsPbBr_3$: $NiO_X$ PNCs. Both kinds of PNCs show 8.5±1.1 nm average size.

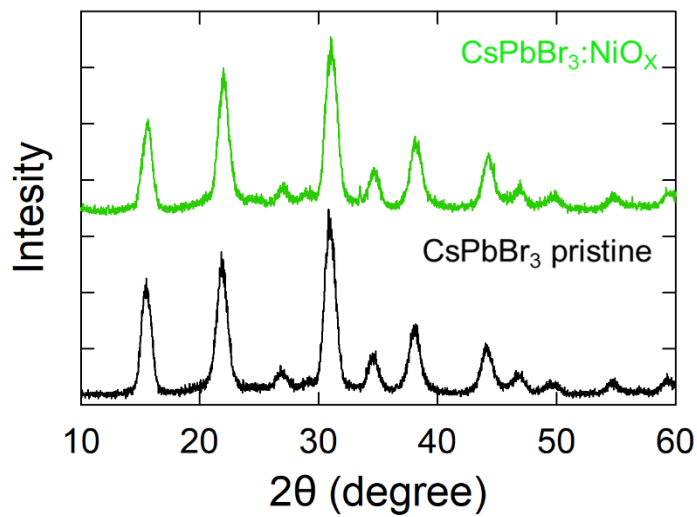

**Figure S14**. X-Ray diffraction patterns of pristine CsPbBr$_3$ PNCs (black curve) and CsPbBr$_3$:NiO$_X$ PNCs (green curve).

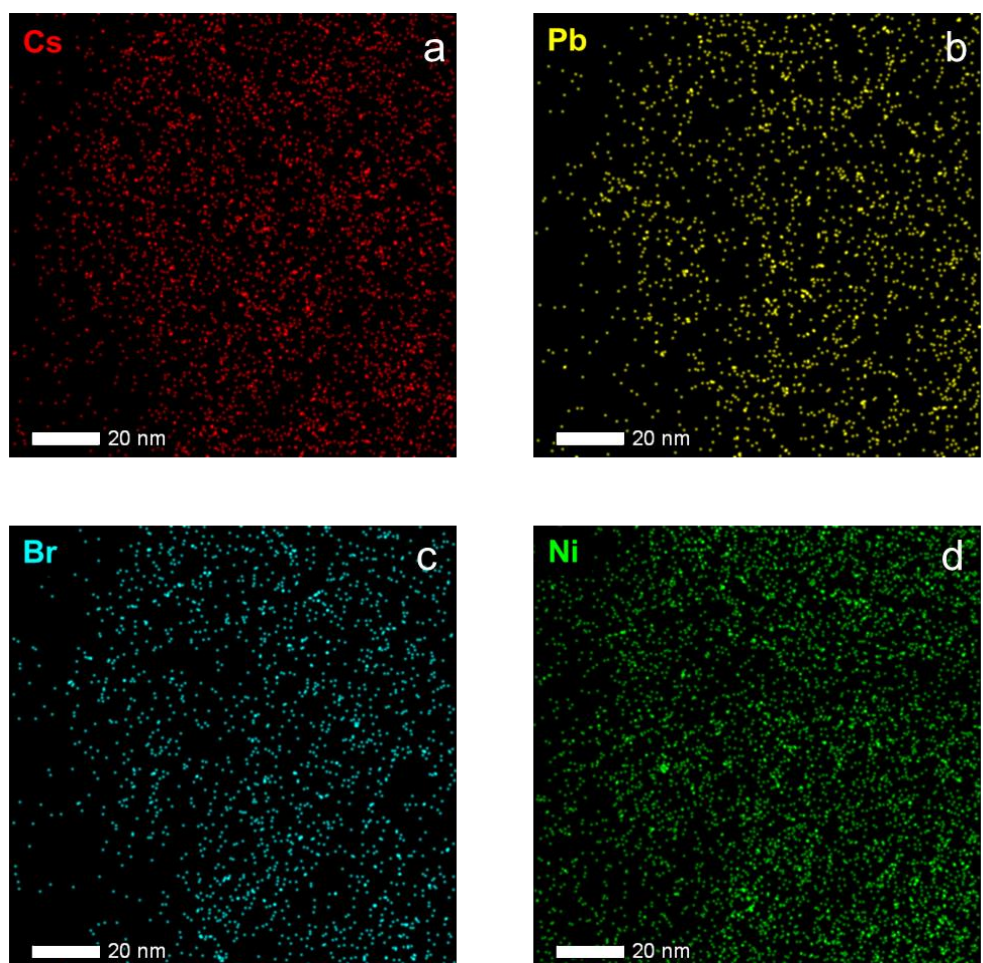

**Figure S15.** Elemental mapping images of Cesium (a), Lead (b), Bromine (c) and Nickel (d) in a CsPbBr$_3$:NiO$_X$ PNCs sample.

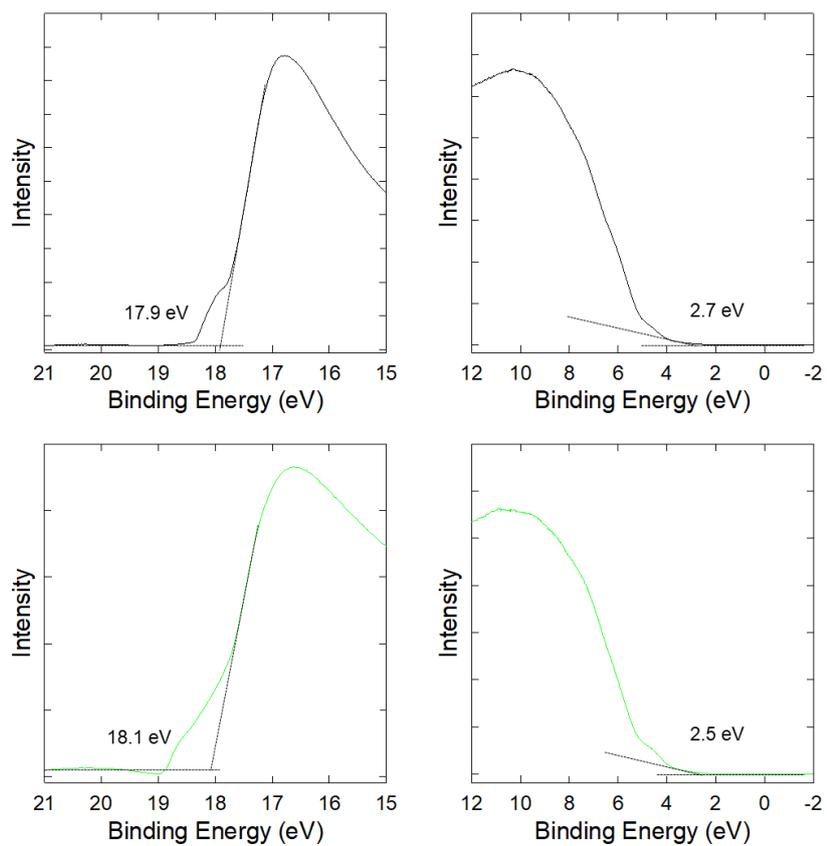

**Figure S16**. UPS spectra of pristine CsPbBr$_3$ (black line, top plots) and CsPbBr$_3$:NiO$_X$ (green line, bottom plots) samples.

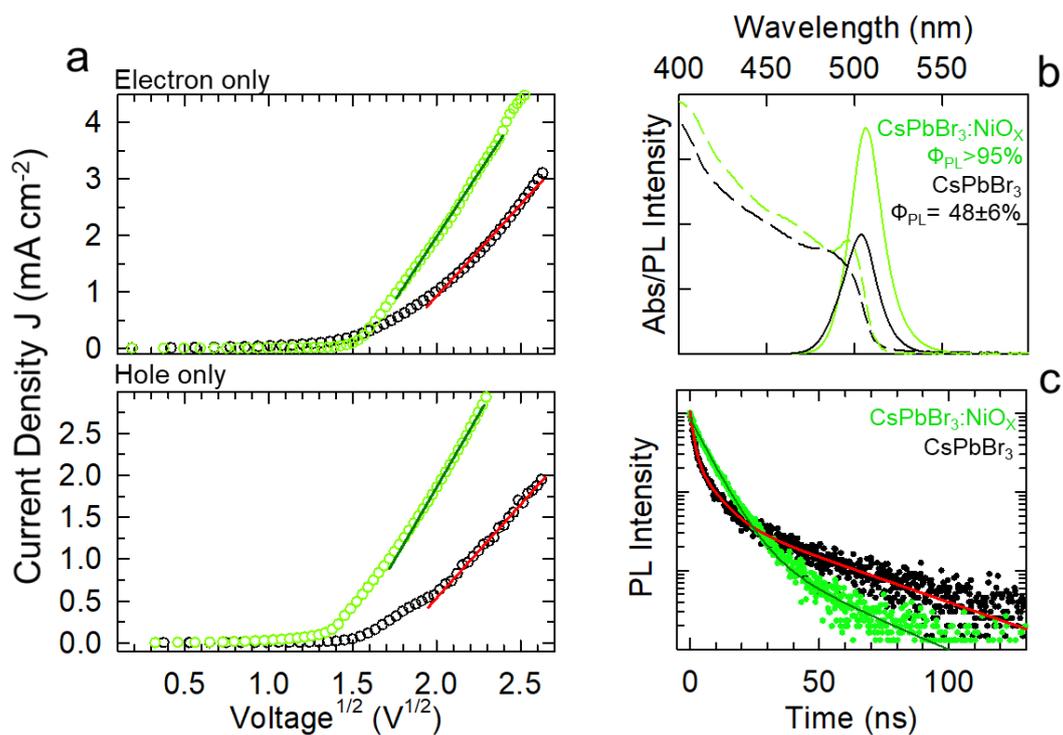

**Figure S17**. a) Carrier mobility (top panel: electrons, lower panel: holes) in single carrier devices embedding pristine $CsPbBr_3$ PNCs (black symbols, or $CsPbBr_3$:$NiO_X$ PNCs (green symbols). Device structures for electron and hole mobility are ITO/ZnO/PNCs/TPBi/LiF/Al and ITO/PEDOT:PSS/Poly-TPD/PNCs/$MoO_3$/Al, respectively. The improvement of the carrier mobilities is likely corroborated by the partial replacement of the oleic acid/oleyl amine ligands with dodecyl dimethylammonium bromide and to the effective passivation of surface carrier traps posited by the $NiO_x$ coating, in agreement with the optical data in panel 'b' and 'c'. b) Optical absorption (dashed lines) and photoluminescence (solid lines) spectra of pristine (black) and $NiO_X$-treated $CsPbBr_3$ PNCs (green) in toluene solution (excitation at 450 nm) and, c) respective photoluminescence decay curves.

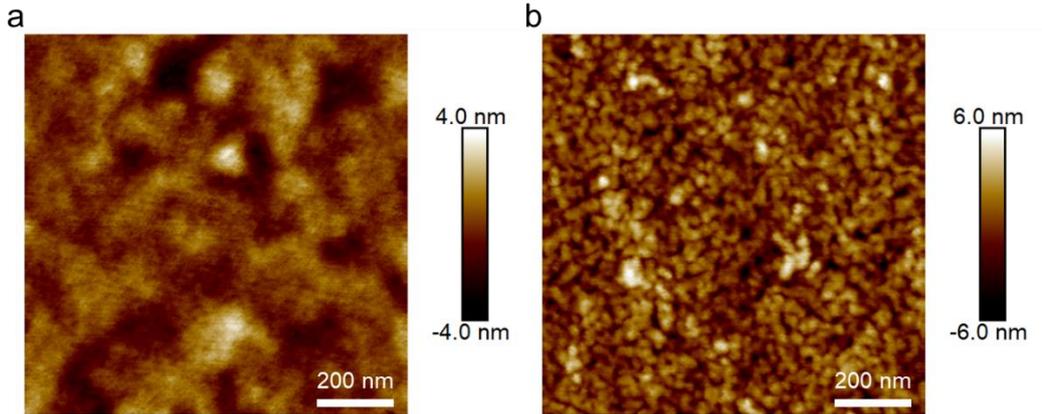

**Figure S18**. AFM images of ITO/PEDOT:PSS/Poly-TPD (a) and ITO/PEDOT:PSS/Poly-TPD/CsPbBr$_3$:NiO$_X$ (b) measured in height mode. The RMS roughness of the films are 1.05 nm (a) and 1.45 nm (b), respectively.

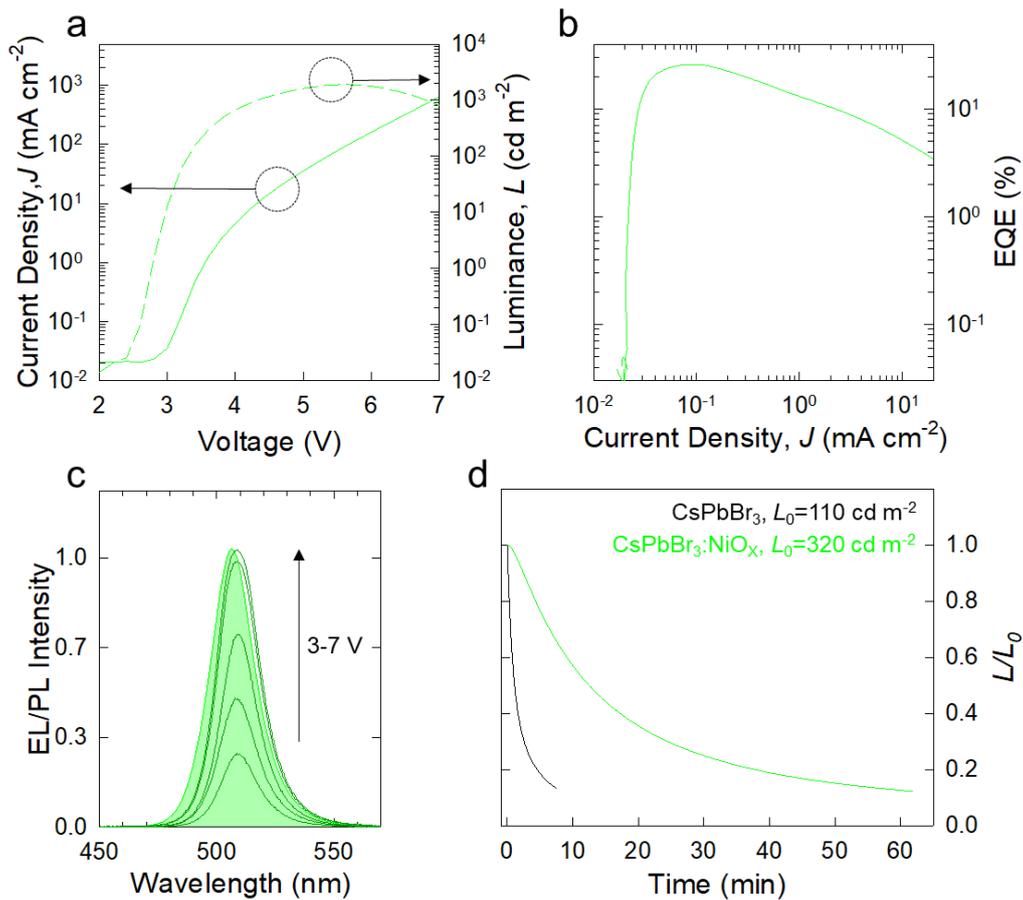

**Figure S19**. a) Current density – Voltage (J-V, continuous line) and Voltage – Luminance (L-V, dashed line) curves of a representative CsPbBr$_3$:NiO$_X$ PNCs-based LED measured at room temperature in air with simple encapsulation. b) EQE curves of the same LEDs reported in 'a'. c) EL spectra (dark green curves) of an ultra-thin LED embedding CsPbBr$_3$:NiO$_X$ ($d$ = 8 nm) at increasing bias voltage (3-7 V as indicated by the arrow) together with the respective PL spectrum (shaded curve, excitation wavelength 450 nm). d) Luminance as a function of operating time of a representative CsPbBr$_3$:NiO$_X$ PNCs-based LED in comparison with a pristine CsPbBr$_3$-based LED prepared with the same synthetic route.

**Table S1:** Electro-optical characteristics of representative LEDs embedding CsPbBr$_3$:NiO$_X$ as a function of emissive layer thickness.

| Emissive layer Thickness | Turn-on Voltage (V) @ 1cd m$^{-2}$ | L$_{max}$ (cd m$^{-2}$) | EQE$_{max}$ (%) |
|---|---|---|---|
| 90 | 5.1 | 188 | 2.4 |
| 39 | 4.4 | 278 | 6.0 |
| 31 | 2.9 | 464 | 8.1 |
| 21 | 2.8 | 1485 | 17.5 |
| 8 | 2.8 | 1900 | 25.5 |

**Table S2:** Electro-optical characteristics of LEDs embedding pristine $CsPbBr_3$ as a function of emissive layer thickness.

| Emissive layer Thickness | Turn-on Voltage (V) @ 1cd m$^{-2}$ | $L_{max}$ (cd m$^{-2}$) | $EQE_{max}$ (%) |
|---|---|---|---|
| 50 | 7.2 | 637 | 0.4 |
| 40 | 5.6 | 573 | 0.6 |
| 20 | 5.6 | 1500 | 1.2 |

**Table S3**. The performance of ultrathin device embedding CsPbBr$_3$:NiO$_x$ NCs synthetized through a triple ligand room temperature method.

| Sample NO. | Turn-on Voltage (V) @ 1 cd m$^{-2}$ | L$_{max}$ (cd m$^{-2}$) | EQE$_{max}$ (%) |
|---|---|---|---|
| 1 | 2.6 | 4162 | 25.7 |
| 2 | 2.6 | 5975 | 21.7 |
| 3 | 2.6 | 6814 | 27.6 |
| 4 | 2.8 | 13810 | 23.1 |
| 5 | 2.6 | 6822 | 23.5 |
| 6 | 2.6 | 6839 | 26.5 |
| 7 | 2.6 | 8702 | 25.4 |
| 8 | 2.6 | 8914 | 21.8 |
| 9 | 2.8 | 10200 | 26.9 |
| 10 | 2.8 | 8322 | 24.3 |
| 11 | 2.8 | 6450 | 22.1 |
| 12 | 2.8 | 9393 | 24.6 |
| 13 | 2.8 | 9187 | 24.5 |
| 14 | 2.8 | 15900 | 22.6 |
| 15 | 2.8 | 13780 | 23.1 |
| 16 | 2.6 | 10240 | 27.1 |
| 17 | 2.6 | 10160 | 25.2 |
| 18 | 2.6 | 12460 | 24.8 |
| 19 | 2.6 | 11520 | 25.2 |
| 20 | 2.6 | 9425 | 24.0 |
| 21 | 2.6 | 8141 | 26.9 |
| 22 | 2.6 | 10120 | 21.6 |
| 23 | 2.8 | 6241 | 24.4 |
| 24 | 2.8 | 5900 | 24.2 |
| 25 | 2.8 | 7243 | 24.2 |
| 26 | 2.8 | 6544 | 23.9 |
| 27 | 2.8 | 5845 | 24.0 |
| 28 | 3.0 | 7575 | 24.5 |
| 29 | 2.8 | 5292 | 22.7 |
| 30 | 2.8 | 5102 | 25.9 |
| 31 | 2.8 | 5579 | 24.8 |

| Sample NO. | Turn-on Voltage (V) @ 1 cd m$^{-2}$ | L$_{max}$ (cd m$^{-2}$) | EQE$_{max}$ (%) |
| --- | --- | --- | --- |
| 32 | 2.8 | 7415 | 27.0 |
| 33 | 2.8 | 6919 | 26.3 |
| 34 | 2.8 | 6229 | 22.9 |
| 35 | 2.8 | 7975 | 27.5 |
| 36 | 2.8 | 8292 | 28.4 |
| 37 | 2.8 | 14940 | 23.0 |
| 38 | 2.6 | 12670 | 22.2 |
| 39 | 2.6 | 11900 | 23.1 |
| 40 | 2.8 | 13100 | 29.2 |

**Table S4**. Performance of ultrathin device embedding CsPbBr$_3$:NiO$_x$ NCs fabricated via conventional hot injection method.

| Sample NO. | Turn-on Voltage (V) @ 1 cd m$^{-2}$ | L$_{max}$ (cd m$^{-2}$) | EQE$_{max}$ (%) |
| --- | --- | --- | --- |
| 1 | 2.8 | 1407 | 25.0 |
| 2 | 2.8 | 1399 | 23.9 |
| 3 | 2.8 | 1491 | 27.4 |
| 4 | 2.8 | 2333 | 24.8 |
| 5 | 2.8 | 1059 | 23.0 |
| 6 | 2.8 | 1050 | 23.3 |
| 7 | 2.8 | 1716 | 21.3 |
| 8 | 2.8 | 1772 | 23.7 |
| 9 | 2.8 | 1125 | 24.3 |
| 10 | 3.0 | 1648 | 22.8 |
| 11 | 2.8 | 1727 | 25.3 |
| 12 | 2.8 | 2132 | 23.2 |
| 13 | 2.8 | 1622 | 25.0 |
| 14 | 2.8 | 1555 | 21.8 |
| 15 | 2.8 | 1775 | 24.8 |
| 16 | 2.8 | 1596 | 22.6 |
| 17 | 2.8 | 1264 | 26.7 |
| 18 | 2.8 | 1918 | 27.2 |
| 19 | 2.8 | 1440 | 23.6 |
| 20 | 2.8 | 1900 | 25.5 |
| 21 | 2.8 | 1474 | 22.5 |
| 22 | 2.8 | 2199 | 21.7 |
| 23 | 2.8 | 1402 | 23.4 |
| 24 | 2.8 | 2529 | 22.9 |
| 25 | 2.8 | 2477 | 25.2 |
| 26 | 2.8 | 3199 | 27.8 |
| 27 | 2.8 | 1230 | 24.3 |
| 28 | 2.6 | 3630 | 25.9 |
| 29 | 2.6 | 2318 | 22.2 |
| 30 | 2.6 | 2598 | 25.7 |
| 31 | 2.6 | 2479 | 21.7 |
| 32 | 2.8 | 3602 | 21.3 |
| 33 | 2.8 | 3436 | 23.2 |
| 34 | 2.6 | 1703 | 27.4 |
| 35 | 2.6 | 2473 | 22.9 |
| 36 | 2.6 | 2385 | 22.2 |
| 37 | 2.6 | 1958 | 21.7 |
| 38 | 2.8 | 1755 | 21.8 |
| 39 | 2.8 | 2003 | 21.2 |
| 40 | 2.6 | 3675 | 26.6 |